\documentstyle[aps,preprint,epsf]{revtex}

\newcommand{\NN}{{\cal N}}
\def\rarr{\rightarrow}
\def\gm{\gamma}
\def\eps{\epsilon}
\def\del{\partial}
\def\bata{scaling coefficient}
\def\susy{supersymmetry}

\def\susic{supersymmetric}

\def\half{{1\over 2}}

\def\rarr{\rightarrow}

\def\NN{{\cal N}}
\def\order{{\cal O}}

\def\none{$\NN=1$}
\def\ntwo{$\NN=2$}

\def\nfour{$\NN=4$}
\def\susy{supersymmetry}
\def\susic{supersymmetric}
\def\hyp{hypermultiplet}

\def\nf{N_f}

\def\support{This research was supported in
        part by National Science Foundation grant NSF PHY-9513835 and
        by the W.M.~Keck Foundation.}
\def\ZZ{{\bf Z}}

\newcommand{\be}[1]{\begin{equation}\label{#1}}
\newcommand{\ee}{\end{equation}}
\newcommand{\FFig}[1]{Fig.~{\ref{fig:#1}}}
\newcommand{\eref}[1]{(\ref{#1})}
\newcommand{\Eref}[1]{Eq.~(\ref{#1})}

\begin{document}
\draft
\pagestyle{empty}
{\tightenlines

\preprint{
\begin{minipage}[t]{3in}
\begin{flushright}
IASSNS--HEP--98/87 \\
hep-ph/9810223 \\
\end{flushright}
\end{minipage}
}

\title{\Large\bf On Renormalization Group Flows \\
and Exactly Marginal Operators  in Three Dimensions}
\author{
{\bf
Matthew J. Strassler}}
\address{\ \\ School of Natural Sciences,
Institute for Advanced Study,
Olden Lane,
Princeton, NJ 08540, USA\\
{\tt strasslr@ias.edu}}
\maketitle
\begin{abstract}
As in two and four dimensions, supersymmetric conformal field theories
in three dimensions can have exactly marginal operators.  These are
illustrated in a number of examples with \nfour\ and \ntwo\
supersymmetry.  The \ntwo\ theory of three chiral multiplets $X,Y,Z$
and superpotential $W=XYZ$ has an exactly marginal operator; \ntwo\
$U(1)$ with one electron, which is mirror to this theory, has one
also.  Many \nfour\ fixed points with superpotentials $W \sim \Phi Q_i
\tilde Q^i$ have exactly marginal deformations consisting of a
combination of $\Phi^2$ and $(Q_i\tilde Q^i)^2$.  However, \nfour\
$U(1)$ with one electron does not; in fact the operator $\Phi^2$ is
marginally irrelevant.  The situation in non-abelian theories is
similar.  The relation of the marginal operators to brane rotations is
briefly discussed; this is particularly simple for self-dual examples
where the precise form of the marginal operator may be guessed using
mirror symmetry.
\end{abstract}
}

%\pacs{???}

\newpage
\pagestyle{plain}
%\narrowtext
%\widetext

\vskip 0.4 in

\draft
\tightenlines

In recent years there has been progress on many fronts in our
understanding of supersymmetric field theories in three and higher
dimensions.  There is a vast literature on conformal field theories in
two dimensions, and superconformal field theories (SCFTs) are well
studied there.  However, SCFTs were not thought common in higher
dimensions until quite recently.  Isolated SCFTs were known in the
1970s for $SU(N)$ gauge theories with $N_f=3N(1-\eps)$ flavors of
matter fields in the fundamental representation.  A long list of
finite theories (continuously infinite sets of fixed points indexed by
a gauge coupling) in four dimensions with \nfour, \ntwo\ and \none\
supersymmetry were discovered in the 1980s (see \cite{emop,emoptalk}
for a list of references) in which a combination of the gauge kinetic
term and an interaction among matter fields serves as an exactly
marginal deformation of the free field theory.\footnote{An ``exactly
marginal operator'' is one which, when added to the Lagrangian of a
CFT, preserves conformal invariance.}  There are also well-known
isolated SCFTs in theories in three dimensions, stemming both from
supersymmetric generalizations of $\phi^4$ theory and from gauge
theories at large $N_f$.  Although it is easy to find exactly marginal
operators in theories based on supersymmetric $\phi^4$ theory, I am
unaware of any research on marginal operators in three-dimensional
theories prior to recent developments.

Seiberg \cite{NAD} showed that SCFTs in four dimensions are much more
common than previously realized.  In \cite{emop}, using techniques
similar to those used in two dimensions but employing also special
properties of four dimensions, it was shown that many of these
non-trivial SCFTs have exactly marginal operators, and that the finite
theories previously studied are just special cases.  After the work of
\cite{nsewddd,kinsddd,ucbbrane,ahew,ucbmirror,ntwobrane,ntwovort,oaahrotate},
our understanding of three dimensional SCFTs greatly increased.
Exactly marginal operators were noted at that time, using some of the
same arguments as in two dimensions
\cite{lancemarg,cringmart,gvw,cringvw,lvw}.  However, nothing was
published on this subject.  This article is intended to fill the gap
in the literature.

While elementary application of the superconformal algebra is often
sufficient to show that a given SCFT contains an operator which is
marginal, it is not sufficient to show that the operator is
{\it exactly} marginal --- that it remains
marginal when it is added to the Lagrangian as a perturbation.
Instead, an argument must be given that the SCFT in question lies
inside a continuous space of SCFTs; motion within this space
corresponds to perturbation by an exactly marginal operator.  In
finite theories in four dimensions, all-orders arguments based on
perturbation theory were given (see \cite{emop,emoptalk} for a list of
references.)  However, these approaches cannot be used if the free
theory is not inside the space of SCFTs under study.  A more powerful
but very simple argument for the existence of continuous sets of SCFTs
is the following. Each beta function is a function of all of the gauge
and superpotential couplings.  Suppose the total number of couplings
is $n$.  Since there is one beta function for each coupling, the
requirement that all beta functions vanish puts $n$ constraints on $n$
couplings; any solution to these constraints is generally isolated and
has no exactly marginal deformations.  However, if only $p$ of the
beta functions are linearly independent as functions of the $n$
couplings, then the general solution to the vanishing of the beta
functions will be an $n-p$ dimensional subspace of the space of
couplings.  Any given SCFT on that subspace will have $n-p$ linearly
independent exactly marginal deformations.  Of course, it is possible
that there are no solutions, or multiple solutions, to the conditions
of vanishing beta functions.

In \cite{emop}, the arguments proving the existence of exactly
marginal operators are based on exact formulas for the beta functions.
These formulas follow from special properties of \none\ (\ntwo)
supersymmetric gauge theories in four (two and three) dimensions.
Such theories have a holomorphic superpotential.  Holomorphy implies
severe restrictions; in particular, couplings of chiral fields in the
superpotential are not perturbatively renormalized
\cite{oldnonren,nonrenthm,powerholo}.  Non-perturbative
renormalizations of the superpotential are restricted by holomorphy
\cite{powerholo,nsexact}. Still, any physical coupling {\it is}
renormalized, and its running can be expressed in terms of its
canonical dimension and the anomalous dimensions of the fields that it
couples.  That is, corresponding to the superpotential
$W=h\phi_1\dots\phi_n$ there is a $\beta$-function
\be{betah} 
\beta_h
\equiv \frac{\del h(\mu)}{\del \ln\mu} = h(\mu)\Big( -d_W+\sum_k
d({\phi_k}) \Big)
= \half h \left[n\left(d-2\right)-2(d-1)
+\sum_{k=1}^n\gm(\phi_k)\right]\equiv \half h(\mu)A_h 
\ee
where $d_W=d-1$ is the canonical dimension of the
superpotential and $d(\phi_k = \half[d-2 +\gamma(\phi_k)]$ is the dimension of the superfield field
$\phi_k$, with $\gamma(\phi_k)$ is its anomalous mass dimension.  I will
refer to $A_h$ as a \bata; it is twice the physical dimension of
the operator $\phi_1\dots\phi_n$.

In four dimensions, exact formulas are also known for the running of
gauge couplings \cite{NSVZbetaA,NSVZbetaB,SVa,SVb}; these formulas
follow from anomalies, and relate the gauge beta functions to
anomalous dimensions of the charged fields in a similar way to
\eref{betah}.  This makes all of the beta functions linear functionals
of the anomalous dimensions, and it is quite easy to find theories in
which the beta functions are linearly dependent.

In three dimensions similar formulas for the gauge couplings have not
yet been found, and it is not clear that they exist.  (It is also
clear that they cannot depend merely on the anomalous dimensions of
the charged fields, as they do in four dimensions; this will be
explained later.)  Despite the absence of such formulas, it is still
possible to demonstrate the existence of exactly marginal operators,
since application of the argument only requires that linear dependence
of two or more of the beta functions be established. For this reason,
our lack of knowledge of the beta function for gauge couplings in
three dimensions is not a hindrance, as long as we consider linear
dependence of beta functions for couplings in the superpotential,
expressed through \Eref{betah}.  In other words, the superpotentials
with exactly marginal operators in three dimensions will have much the
same form as in two dimensions, and will not have the more general
form possible in four dimensions where linear dependence of gauge beta
functions may be included.  On the other hand, gauge interactions
still play an essential role in three dimensions by expanding the
number of possible exactly marginal operators, as will be explained
below.  In this sense, the results described in this letter are
intermediate between those of two and four dimensions.

%\section{Theories Without Gauge Interactions}

The simplest \ntwo\ superconformal field theory (SCFT) is the
supersymmetric generalization of $\phi^4$ theory.  The $\lambda\phi^4$
perturbation of a free scalar field $\phi$ is relevant in three
dimensions, and flows to a well-studied fixed point.  The perturbation
$W=\lambda\Phi^3$ of a free chiral \ntwo\ superfield $\Phi$ is similarly
relevant, since $\lambda$ has mass dimension $+\half$ at the free
field theory.  The beta function for $\lambda$ is exactly
\be{betalam}
\beta_\lambda = \lambda (-2 + 3 d_\phi) = \half\lambda (-1 + 3 \gamma_\phi)
\ee
($\lambda$ has mass dimension $+\half$ at the free field theory) and
drives the dimension of $\Phi$ upward.
\be{dphi}
d_\Phi(\lambda) = \half + \order(\lambda^2) > d_\Phi(0)
\ee
This growth continues (presumably monotonically, though no techniques
as yet can prove it) until $d_\Phi=2/3$.  At this point $\lambda$ is
dimensionless and its beta function vanishes.   It is to be expected that
this fixed point is stable; the beta function is negative (positive)
if $\lambda$ is smaller (larger) than its fixed point value as long as
$d_\Phi$ passes monotonically through $2/3$ in the vicinity of the
fixed point.  Since it would require fine tuning for $d_\Phi(\lambda)$
to reach a maximum of $2/3$ precisely at the fixed point value of
$\lambda$, it is almost certain that the fixed point is
stable. Stability can also be checked in an epsilon expansion,
although this does not preserve supersymmetry.

\begin{figure}
\centering
\epsfxsize=5.0in
\hspace*{0in}\vspace*{0.2in}
\epsffile{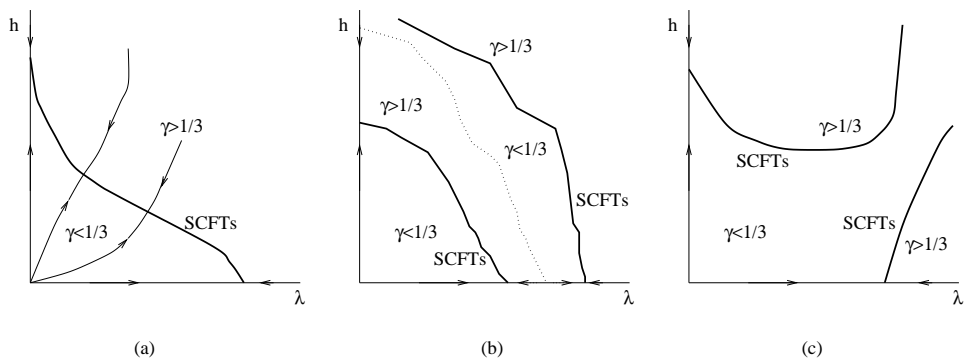}
\caption{The theory $W=\lambda(X^3+Y^3+Z^3)+hXYZ$ has one (a) or more
(b,c) one-complex-dimensional spaces of SCFTs, which may (a,c) or may
not (b) connect the SCFTs with $\lambda=0$ and $h=0$.  The SCFTs separate regions
where $\gamma>1/3$ from those with $\gamma<1/3.$ Renormalization group
flow toward the infrared is indicated by arrows.  The dotted line in (b)
indicates a line of infrared unstable SCFTs.}
\label{fig:XYZ}
\end{figure}

This SCFT can be used to create a theory with a exactly marginal
operator. Consider three chiral superfields $X,Y,Z$ with superpotential
$W=\lambda_X X^3+\lambda_Y Y^3+\lambda_Z Z^3$. 
At the fixed point, $\lambda_X =\lambda_Y
=\lambda_Z$ and $d_X = d_Y = d_Z = 2/3$.  The perturbation  $\Delta W=hXYZ$
leads to an exactly marginal operator:
\be{XYZop}
A_{\lambda_X} = -1+3\gamma_X \ ; \
A_{\lambda_Y} = -1+3\gamma_Y \ ; \
A_{\lambda_Z} = -1+3\gamma_Z \ ; \
A_{\lambda_H} = -1 + \gamma_X +\gamma_Y + \gamma_Z \ .
\ee
Only three of these \bata s are linearly independent, so there is one
exactly marginal operator, lying by symmetry in the subspace
$\lambda_X=\lambda_Y=\lambda_Z\equiv \lambda_0$ (where also
$\gamma_X=\gamma_Y=\gamma_Z)$ defined by the condition that all \bata
s vanish, namely the single constraint on two couplings
$\gm_X(\lambda_0,h)=1/3$.  (Note that we could have assumed the
symmetry among $X,Y,Z$ from the beginning and arrived at the same
number of marginal operators; in future I will often shorten the
analysis by making analogous assumptions.)  Within the two-dimensional
complex space of couplings $\lambda_0$ versus $h$, there will be a
one-complex-dimensional subspace, separating the regions of
$\gm_X<1/3$ and $\gm_X>1/3$, on which the theory is conformal, as
shown in \FFig{XYZ}. The SCFTs $ W=\lambda(X^3 +Y^3+Z^3)$ and $W=hXYZ$
are thus connected by a line of SCFTs, as shown in \FFig{XYZ}a. It is
also possible there are multiple subspaces, as shown in \FFig{XYZ}b and
\FFig{XYZ}c, in which case the last statement may or may not be true.
%%%%%%%%%%%%%%%%%%%%%%%%
%%%%%%%%%%%%%%%%%%%%%%%%% 

Note that the perturbations $X^2Y,Y^2Z$, and so forth are
not marginal deformations of the theory with $W=\lambda(X^3+Y^3+Z^3)$.
These perturbations are redundant, as they can be removed by
field redefinitions.  The only non-redundant perturbation is
the operator $XYZ$.  Similar statements apply to generalizations
of this model, such as
$W= \sum_{i=1}^6 \lambda X_i^3$ which has exactly marginal operators
$X_1X_2X_3+X_4X_5X_6$, $X_2X_3X_4+X_5X_6X_1$, {\it etc.}

No other interesting SCFTs can be built using chiral superfields
without gauge interactions.  The perturbation $\Delta W=hX_1X_2X_3X_4$
of any SCFT (including a free theory) cannot be exactly marginal, on
general grounds following from the conformal algebra.  The dimension
of at least one of the fields $X_i$ must be $1/2$ or less for
$X_1X_2X_3X_4$ to be marginal.  However, at a fixed point no gauge
invariant field can have dimension less than $1/2$ (and therefore all
four fields must have dimension $1/2$ for $X_1X_2X_3X_4$ to be
marginal) and fields of dimension $1/2$ are free (and therefore $h$
cannot be non-zero at a fixed point.)  Thus these perturbations must
be marginally irrelevant.  Note that the same reasoning shows that
$\phi^6$ and $\phi^2\bar\psi\psi$ perturbations of non-\susic\
theories of free scalars and fermions are irrelevant.

%\section{Theories with Abelian gauge symmetries}

Since the above limitation stems from having all fields gauge
invariant, it is natural as a next step to introduce gauge
interactions in hope of evading it.  The simplest theories
to study are \ntwo\ and \nfour\ \susic\ theories
with a $U(1)$ gauge group and some charged matter.

It is useful to review the properties of \nfour\ gauge theories in
the language of  \ntwo\ \susy.  In \ntwo\ language, the
\nfour\ theory of $U(1)$ with $N_f$ hypermultiplets of charge 1 has
a $U(1)$ vector multiplet, a neutral chiral multiplet $\Phi$, chiral
multiplets $Q_i,\tilde Q^i$ of charge $1$ and $-1$, and a superpotential
$W=\Phi \sum_i Q_i\tilde Q^i$.   At the origin of
moduli space this theory flows in the infrared to an SCFT.  While
symmetry tells us $d_Q=d_{\tilde Q}$, and \ntwo\ \susy\ tells us
through the dimension of the superpotential that $d_\Phi
+ 2d_Q = 2$, all of \nfour\ \susy\ is required to conclude that
$d_\Phi=1$ and $d_Q=\half$, the latter being its canonical dimension.
In short, only $\Phi$ picks up an anomalous dimension.  The shift in
dimension of $\Phi$ from $\half$ to 1 is directly linked to the shift
of the gauge coupling $g$ from dimension $\half$ to $0$; the dimension
of $g\Phi$ is constant.  Without \nfour\ \susy, the field $\Phi$ would
not be tied to the gauge boson by any symmetry and there would be no
direct connection between the dimension of $g$ and the dimension of
$\Phi$.  These statements --- that $\Phi$ has dimension $1$ and $Q$
dimension $\half$ --- are true in {\it all} \nfour\ SCFTs which stem
from local Lagrangians, including theories with multiple and/or
non-abelian gauge groups.	

As an aside, notice that this implies that the beta function for the
gauge coupling cannot simply be proportional to a linear combination
of the beta functions of charged fields, as it is in four dimensions.
In the \nfour\ $U(1)$ theories, the anomalous dimensions of $Q,\tilde
Q$ are zero, both in the free theory and at the SCFT, and the fields
$\Phi$ are neutral.  Since the beta function is non-zero in the
ultraviolet and vanishes in the infrared, it must get a non-trivial
contribution from sources other than the charged fields.

Mirror symmetry \cite{kinsddd}, a relation between two SCFTs, will
also be useful to us in the following.  Under mirror symmetry, the
SCFT of \nfour\ $U(1)$ with one hypermultiplet is mapped to a theory
of a single free hypermultiplet $q,\tilde q$ \cite{nsewddd}.  The
(nonlocal) vortex creation operators $V_\pm$ of the original theory
are the fields $q$ and $\tilde q$ \cite{ntwovort}, while the operator
$\Phi$ is mapped to $q\tilde q$ \cite{kinsddd}.  (Note that this
requires $V_+V_-\propto\Phi$, which is indeed satisfied in the
original theory.)  The free mirror theory has perturbations
$(q)^m(\tilde q)^n$ which are relevant for $m+n<4$, marginally
irrelevant for $m+n=4$, and irrelevant for $m+n>4$.  Those with $m\neq
n$ are hard to interpret in the original theory, as they involve the
poorly understood operators $V_\pm$.  However, we may make unambiguous
statements about the perturbations $\Phi^k$, which break \nfour\
\susy\ to \ntwo\ for $k>1$.  Only the perturbation $k=1$ --- a
Fayet-Iliopolous term --- is relevant; it corresponds to a mass for
the mirror hypermultiplet \cite{kinsddd}.  The perturbation $\Phi^2$,
which is mapped to $(q\tilde q)^2$, is marginally irrelevant.  This is
not the usual expectation for a mass term.  In fact, a non-zero mass
for $\Phi$ is relevant, and drives the theory away from \nfour\ \susy,
at any finite distance from the SCFT, since $\gm(\Phi)<1$ until the
fixed point is precisely reached.  Only at the fixed point itself is
the mass term marginal, and more specifically marginally irrelevant.
The meaning of this in a wider context will be discussed shortly.

Next, let us consider \ntwo\ $U(1)$ with one flavor, the same as the
previous theory but with the field $\Phi$ removed and a superpotential
$W=0$.  As shown in \cite{ntwobrane,ntwovort}, this theory is mirror
to a theory of three singlets $S,q,\tilde q$ with superpotential
$W=Sq\tilde q$.  This is precisely of the form $W=XYZ$, a theory
considered earlier and shown to have an exactly marginal operator when
$X^3+Y^3+Z^3$ is added as a perturbation.  The mapping of operators is
$Q\tilde Q, V_+,V_-\rarr S,q,\tilde q$.  From the discussion of
the $W=XYZ$ model we learn that $Q\tilde Q$ has dimension $2/3$, and thus
$d_Q=d_{\tilde Q} = 1/3$.  This means that the operator $(Q\tilde
Q)^2$ is relevant, as are $V_+^2$ and $V_-^2$, while $(Q\tilde Q)^3 +
V_+^3+V_-^3$ is an exactly marginal operator corresponding to the
mirror of $S^3+q^3+\tilde q^3$.

It is interesting to consider the relevant operator $(Q\tilde Q)^2$ in
this theory.  The importance of this operator was first emphasized in
\cite{emop}, where it was shown that in four dimensions it connects
the electric-magnetic duality of finite \ntwo\ theories to that of
self-dual \none\ theories.  It has also been considered in
\cite{kinstwo,ntwobrane,ntwovort,oaahrotate,ikew}.  Here, it plays a similar
interesting role.  As in \cite{emop,kinstwo,ntwobrane,ntwovort} one
may rewrite the superpotential $W=\half hQ\tilde QQ\tilde Q$ by
introducing a gauge singlet auxiliary field $M$ with superpotential
$W=MQ\tilde Q -{1\over 2h}M^2$.  Although $M$ is introduced as an
auxiliary field, it develops a propagator through loop diagrams and is
indistinguishable from a canonical field as far as infrared physics is
concerned.  Thus, although the theory with the new field is not the
same as the original one, it has the same infrared behavior.  I claim
that the physical value of $h$ flows to infinity in the infrared;
consequently the mass of $M$ goes to zero, and we are left in the
infrared with the \nfour\ SCFT theory of $U(1)$ with one flavor
considered just above.  To check this claim, consider the
mirror description: the field $S$ becomes massive, leading to a
quartic superpotential
\be{Ssqrd}
W=Sq\tilde q + \half hS^2 \rarr -{1\over2h}(q\tilde q)^2 
\ee
which is marginally irrelevant.  The fields $q,\tilde q$ are
thus free in the infrared, and constitute a free \nfour\ \susic\
\hyp.

\begin{figure}
\centering
\epsfxsize=3.0in
\hspace*{0in}\vspace*{0.2in}
\epsffile{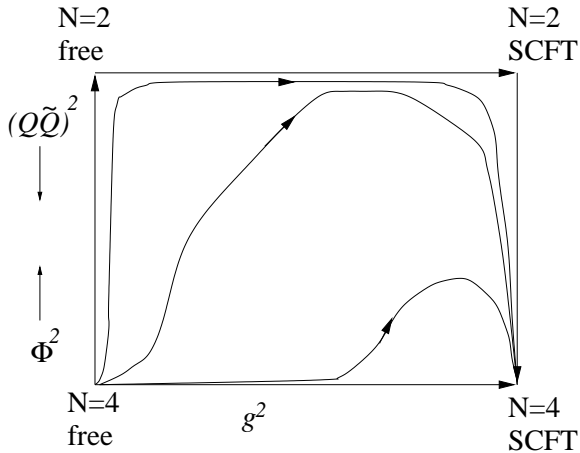}
\caption{Renormalization group flow connecting the \nfour\ and \ntwo\
theories of $U(1)$ with $\nf=1$.  The horizontal direction represents
the gauge coupling; the vertical represents the coupling of $\Phi^2$
(which grows as one moves up the diagram) or equivalently the
coupling of $(Q\tilde Q)^2$ (which grows as one moves down the diagram.)}
\label{fig:oneflavor}
\end{figure}

From this we learn more about the global behavior of the
renormalization group flow connecting these two theories.  If we begin
with the free \nfour\ theory in the ultraviolet, the perturbation $\half
m\Phi^2$ is relevant. The theory flows toward the free \ntwo\ theory.
However, it does so with an operator $(Q\tilde Q)^2$ in the
superpotential.  If $m\gg g^2$ then a classical analysis applies, and
the operator $(Q\tilde Q)^2$ is just barely relevant; the
theory flows very close to the \ntwo\ free theory, then very close to
the \ntwo\ SCFT, then away from the \ntwo\ fixed point toward
the \nfour\ SCFT.  On the other hand, if $m\ll g^2$, then the
gauge coupling almost reaches its \nfour\ fixed point before the
effect of $m\neq 0$ drives the theory away from the \nfour\ \susic\
theory.  However, it cannot get too far, as the $(Q\tilde Q)^2$
operator rapidly becomes relevant, driving the theory back to the
\nfour\ SCFT.  The flow is schematically shown in
\FFig{oneflavor}.

%%%%%%%%%%%%%%%%%%%%%%%%
%%%%%%%%%%%%%%%%%%%%%%%%% 
Now let us consider $N_f>1$, beginning with the \ntwo\ case.  These
theories have $W=0$ and flow to fixed points whose mirror descriptions
have gauge groups $U(1)^{\nf-1}$ and $\nf$ triplets of mirror matter
fields $S_i,q_i,\tilde q_i$ (the last two having charges under the
$i^{th}$ and $(i-1)^{th}$ gauge groups) with cubic superpotentials
$W=S_iq_i\tilde q_i$.  However, no symmetry relates $S_i$ to $q_i$, in
contrast to the case for $\nf=1$.  Therefore, the anomalous dimensions
of these fields cannot be determined.  In the absence of any
superpotential in the original variables, which might permit the use
of \Eref{betah}, and in the absence of knowledge of the low-energy
anomalous dimensions, it is impossible at this time to determine
whether any of these \ntwo\ fixed points have exactly marginal
operators.

However, the fields $S_i$, which under mirror symmetry are mapped to
linear combinations of $Q_i\tilde Q^i$, are unlikely to have dimension
greater than $1$.  We have seen that for $\nf=1$ the dimension of
$Q\tilde Q$ is $2/3$, while for large $\nf$ one can show the dimension
of $Q_i\tilde Q^i$ is less than 1 by an effect of order $1/\nf$.  This
strongly suggests that bilinear terms in $S_i$ are always relevant,
and thus quartic terms $Q_i\tilde Q^jQ_m\tilde Q^n$ are relevant
perturbations of the low-energy fixed point which then cause the
theory to flow, possibly to a new \ntwo\ SCFT.  Let us therefore
consider the fixed points of $U(1)$ with $\nf>1$ and the
superpotential
\be{WQQQQ}
 W = \sum_{k=0}^{[\nf/2]-1} 
\half y_k\left(\sum_{n=1}^{\nf} e^{2\pi i kn/\nf} Q_n\tilde Q^n\right)
\left(\sum_{n=1}^{\nf} e^{-2\pi i kn/\nf} Q_n\tilde Q^n\right)
\ee
(Here $[\nf/2]$ means the integer part of $\nf/2$.) 
This is by no means the most general quartic superpotential, but it
will serve to illustrate some important points.

First, the superpotential preserves a $Z_N$ symmetry relating the
$Q_n$ and $\tilde Q_n$ to one another, and so they all have the same
anomalous dimension $\gm_Q(y_k,g)$.  This is {\it essential} to ensure
that each coupling $y_k$ in \eref{WQQQQ} does not break up into
multiple couplings under the renormalization group flow; if $Q_1$ and
$Q_2$ have different anomalous dimensions, then the couplings
multiplying $(Q_1\tilde Q^1)^2$ and $(Q_2\tilde Q^2)^2$ will run
differently.  With this symmetry, all couplings run with
$A_{y_k}\propto 4\gamma_Q(y_k,g)$ except the gauge coupling, which has
a zero at some point $y_k=0,g=g_0^*$.  If there is some point
$g\neq0$, $y_k\neq0$, where $\gamma_Q(g,y_k)=0$ and $\beta_g(g,y_k)=0$ (two
constraints on $[\nf/2]+1$ couplings) then there will be a space of
SCFTs of complex dimension $[\nf/2]-1$ passing through that point.  

\begin{figure}
\centering
\epsfxsize=3.0in
\hspace*{0in}\vspace*{0.2in}
\epsffile{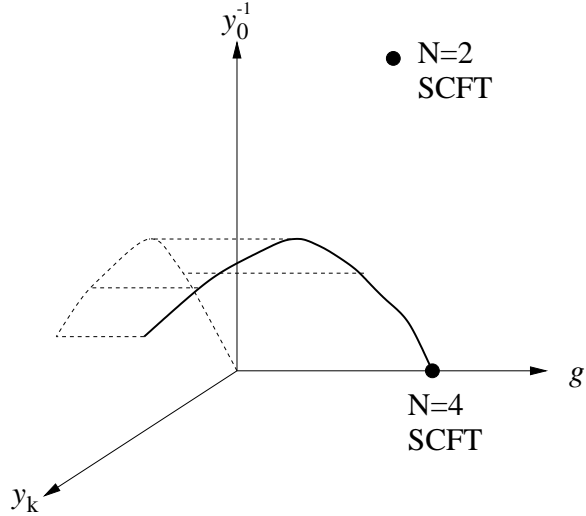}
\caption{For $\nf>1$, the $U(1)$ \nfour\ fixed point, at $y_0=\infty,
y_k=0$, has exactly marginal deformations which preserve \ntwo\ \susy.
In the same space, the \ntwo\ theory with $W=0$ is an isolated SCFT.
There could be more SCFTs than shown.}
\label{fig:manyflavors}
\end{figure}

The limit $y_k\rarr 0$ for $k>0$ and $y_0\rarr \infty$ is a special
one.  As discussed earlier we may make the replacement
\be{replace}
 \half y_0 \left(\sum_{n=1}^{\nf}Q_n\tilde Q^n\right )^2
\rarr W = - {1\over 2 y_0} \Phi^2
+ \Phi  \sum_{n=1}^{\nf}Q_n\tilde Q^n
\ee
so that in the above limit we might obtain the \nfour\ theory of
$U(1)$ with $\nf$ hypermultiplets.  Since we know the \nfour\ theory
has a fixed point with $\gamma_Q=0$, we learn that the theory
\eref{WQQQQ} {\it does} have an $[\nf/2]-1$ dimensional space of fixed points,
and that it contains the \nfour\ SCFT.  The physical picture as a
function of the gauge coupling $g$, the coupling $y_0$, and the other
couplings $y_k$ (treated as a single axis) is shown in
\FFig{manyflavors}.
%%%%%%%%%%%%%%%%%%%%%%%%
%%%%%%%%%%%%%%%%%%%%%%%%% 

The same result may be obtained from the mirror of the theory in
\Eref{WQQQQ}.  The quartic terms in the $Q_i$ correspond to quadratic
terms in the fields $S_i$, which in turn lead to quartic terms in the
mirror fields $q_i\tilde q^i$.  An analysis of these terms leads to
the same conclusion concerning the number of marginal operators, and
also demonstates that the \nfour\ SCFT is present in the limit
$y_0\rarr\infty$ with the other $y_k=0$.

This analysis may be easily generalized to theories with more abelian
gauge groups, and also for theories with non-abelian gauge groups.
The conclusion is the same.  If a \nfour\ SCFT has only a Coulomb
branch, and thus is mirror to a theory of free \hyp s with no gauge
fields, then it has no exactly marginal operators.  Otherwise, the
marginal masses for the fields $\Phi_n$ and the marginal quartic terms
in the hypermultiplets $Q_i$ can generally be balanced off against the
original superpotential $\Phi_n Q_i\tilde Q^i$ to make exactly
marginal operators.

\begin{figure}
\centering
\epsfxsize=3.0in
\hspace*{0in}\vspace*{0.2in}
\epsffile{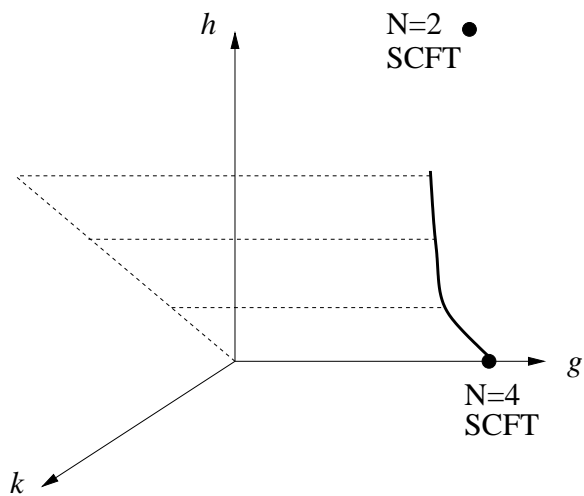}
\caption{In self-dual theories the line of SCFTs lies at $h=k$.}
\label{fig:selfdual}
\end{figure}

A special case with more interesting structure involves those theories
which are self-dual under mirror symmetry.\footnote{Special properties
of quartic superpotentials in self-dual theories in four dimensions
were studied in \cite{emop}; those discussed here are similar but not
identical.}  I use the simplest case, $U(1)$ with $\nf=2$, for
illustration. 
The mirror superpotential is
\be{mirrorW}
W=\phi(q_1\tilde q^1 + q_2\tilde q^2)
\ee
Mirror symmetry maps operators in the following way:
\be{mirrormap}
\Phi \leftrightarrow (q_1\tilde q^1- q_2\tilde q^2)\ ; \
Q_1\tilde Q^1 - Q_2\tilde Q^2\leftrightarrow \phi
\ee
This means that the superpotential
\be{sdmarg} W = \Phi \sum_{n=1}^2 Q_n\tilde Q^n+ \half h \Phi^2 +
\half k (Q_1\tilde Q^1 - Q_2\tilde Q^2)^2 = - {1\over 2h} (Q_1\tilde
Q^1 + Q_2\tilde Q^2) + {k\over 2} (Q_1\tilde Q^1 - Q_2\tilde Q^2)^2
\ee
is mirror to a theory with
\be{sdmargmir}
W = \phi(q_1\tilde q^1+q_2\tilde q^2) + 
\half h (q_1\tilde q^1-q_2\tilde q^2)^2 + 
\half k  \phi^2
={h\over 2} (q_1\tilde q^1-q_2\tilde q^2)^2 -
{1\over 2k}(q_1\tilde q^1+q_2\tilde q^2)^2
\ee
Self-duality is maintained for $h=k$, and the line of SCFTs will lie
along this line by symmetry.  This is shown if \FFig{selfdual}.  In
the limit $h,k\rarr \infty$, by introducing
the auxiliary scalar as in \Eref{Ssqrd}, we obtain the same SCFT as
for $h=k=0$ (the sign in the superpotential can be removed
by a field redefinition.)
%%%%%%%%%%%%%%%%%%%%%%%%
%%%%%%%%%%%%%%%%%%%%%%%%% 

\begin{figure}
\centering
\epsfxsize=3.0in
\hspace*{0in}\vspace*{0.2in}
\epsffile{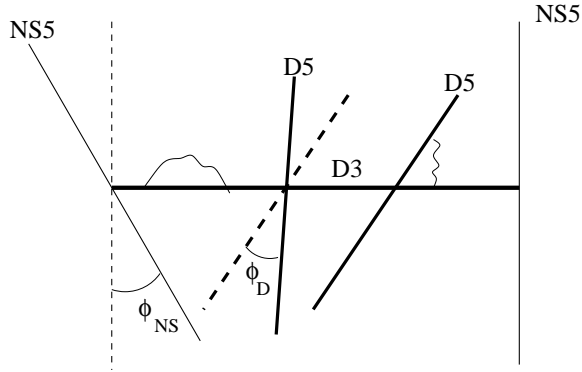}
\caption{Brane construction of $U(1)$ with $\nf=2$ out of
a D3 brane stretched between two NS branes with two D5 branes
placed along it.  Strings with both ends on the D3 are in a vector
multiplet; strings with ends on D3 and D5 are in a hypermultiplet.
Rotations of the D5 and NS5 branes break \nfour\ \susy\ to \ntwo\
and generate the $\Phi^2$ and $(Q\tilde Q)^2$ terms required for
the exactly marginal operators.  In this self-dual case,
the exactly marginal operator preserves self-duality and lies at 
$\phi_{NS}=\phi_D$, as shown.}
\label{fig:branemake}
\end{figure}

As shown in \cite{ucbbrane,ahew} in the context of Type IIB string
theory, the field theory of $U(1)$ with $\nf$ \hyp s can be
constructed by suspending a D3 brane between two NS5 branes, which
gives a $U(1)$ gauge theory whose photon is a 3-3 string (a string
with both ends attached to the D3 brane,) and placing across it $\nf$
D5 branes, which gives 5-3 strings which are \hyp s charged under the
$U(1)$ gauge therory (\FFig{branemake}). All of the branes fill three
dimensions; the NS5 branes also fill dimensions $x^3,x^4,x^5$, the D5
branes fill dimensions $x^7,x^8,x^9$, and the D3 branes stretch across
$x^6$.  We may consider rotating NS5 branes or D5 branes in the
$(x^4,x^5)-(x^8,x^9)$ plane.  This breaks \nfour\ \susy\ to \ntwo.  As
shown in \cite{barbrotate,oaahrotate}, an angle between NS5 branes leads
to a mass term for the field $\Phi$, while rotation of a D5 brane
changes the coupling of its hypermultiplet to $\Phi$.  It is easy to
see that if both D5 and NS5 rotations are considered, and the field
$\Phi$ is integrated out, the rotations correspond to varying the
couplings of quartic terms in the fields $Q,\tilde Q$, as in
\Eref{WQQQQ}.  While a complete and detailed analysis will not be
performed here, it is easy to see that in the case $\nf=2$, the angle
between the two NS5 branes corresponds to $h$ and that between the two
D5 branes corresponds to $k$ in \Eref{sdmarg}.  Clearly self-duality
is maintained only if the angles are identical --- in short, if $h=k$
--- so that the brane construction remains invariant under exchange of
NS5 and D5 branes.  Note that the classical brane construction gives
incomplete insight, however, into the issue of whether the rotation is
exactly marginal.  This is especially obvious in the case of $\nf=1$
where the relative rotation of the NS5 branes is relevant away
from the fixed point but marginally irrelevant in the SCFT.  It would
be nice if this could be understood using a quantum mechanical
treatment of the branes.
%%%%%%%%%%%%%%%%%%%%%%%%
%%%%%%%%%%%%%%%%%%%%%%%%% 

\begin{figure}
\centering
\epsfxsize=3.0in
\hspace*{0in}\vspace*{0.2in}
\epsffile{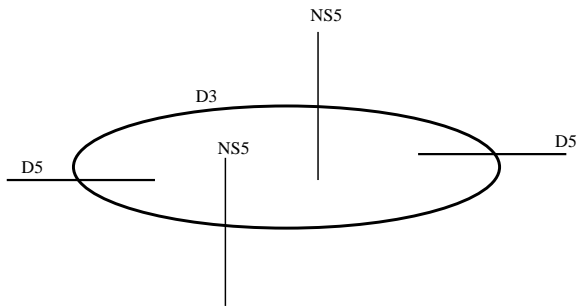}
\caption{A self-dual elliptic model; here the number of
D3 branes is arbitrary.}
\label{fig:elliptic}
\end{figure}

It is also interesting to consider the elliptic models of \cite{ahew},
the simplest being the self-mirror theory shown in \FFig{elliptic}.
Again, if the two NS5 branes and two D5 branes have equal relative
angles, the theory will remain self-dual and the angle will correspond
to an exactly marginal operator.  Under T-duality this theory
corresponds to D2 branes moving on an $R^4/\ZZ_2$ orbifold with two D6
branes, or in M theory to M2 branes on an $R^4/\ZZ_2\times R^4/ \ZZ_2$
orbifold.  Rotations of both types of branes corresponds to deforming
both $\ZZ_2$ orbifolds toward conifold singularities; if the
deformation preserves the $\ZZ_2$ which exchanges them, then
self-duality is preserved and the deformation is marginal.
%%%%%%%%%%%%%%%%%%%%%%%%
%%%%%%%%%%%%%%%%%%%%%%%%% 

These statements can also be generalized to the non-abelian case.  The
precise field theory mapping between operators under mirror symmetry
has not yet been carried out, although it is clearly very similar to
the abelian case.  It would be interesting to understand the
 exactly marginal deformations of the \nfour\ theories in the limit of many
D3 branes, where the AdS/CFT correspondence can be used.  The example
in \FFig{elliptic} and its deformation by brane rotations
\cite{ohtatar} is similar to the example of the $\ZZ_2$ orbifold
deformed to the conifold \cite{ikew}, which corresponds to the same
model with the D5 branes removed.  In four dimensions, the rotation of
the NS5 branes is relevant, and the theory flows to a new fixed point
in which rotation of the NS5 branes is exactly marginal.  In the three
dimensional case, the classical deformations are similar but the
associated dynamics are quite different.  It would be useful to
understand how this is manifested in the supergravity description of
these theories.

\

I thank J. Bagger, K. Intriligator, A. Kapustin,
N. Seiberg, R. Tatar and A. Uranga for discussions.  I especially
thank R. Tatar for encouraging me to publish this work and for sharing
preliminary versions of his own results with K. Oh.  \support
   
%  \nocite{*}                %this uses *everything* in the .bib file
   \bibliography{3dmarg}        %or whatever your .bib file is
\bibliographystyle{utphys}   %if you use utphys.bst

\end{document}